# Children and the Data Cycle: Rights and Ethics in a Big Data World[1]

Gabrielle Berman
UNICEF Office of Research, Innocenti
Florence, Italy
gberman@unicef.org

Kerry Albright
UNICEF Office of Research, Innocenti
Florence, Italy
kalbright@unicef.org

## ABSTRACT

In an era of increasing dependence on data science and big data, the voices of one set of major stakeholders – the world's children and those who advocate on their behalf – have been largely absent. A recent paper estimates one in three global internet users is a child, yet there has been little rigorous debate or understanding of how to adapt traditional, offline ethical standards for research, involving data collection from children, to a big data, online environment (Livingstone et al., 2015). This paper argues that due to the potential for severe, long-lasting and differential impacts on children, child rights need to be firmly integrated onto the agendas of global debates about ethics and data science. The authors outline their rationale for a greater focus on child rights and ethics in data science and suggest steps to move forward, focussing on the various actors within the data chain including data generators, collectors, analysts and end users. It concludes by calling for a much stronger appreciation of the links between child rights, ethics and data science disciplines and for enhanced discourse between stakeholders in the data chain and those responsible for upholding the rights of children globally.

## 1. INTRODUCTION

UNICEF has a specific mandate to protect, respect and uphold the rights of children and their families globally and to help facilitate the full implementation of the Convention on the Rights of the Child (CRC) (UN General Assembly 1989). In undertaking research, and particularly research involving children, that mandate is clear with well-defined guidance provided by international initiatives such as the Ethical Research Involving Children programme (Graham et al., 2013). However, less international attention has been given to rigorous international frameworks for children's data collection and analysis. UNICEF has developed a mandatory cross-organizational procedure on ethical evidence generation (UNICEF, 2015) underpinned by a belief that ethical principles and a rights-based approach are not only relevant in research, but are equally important within all forms of data collection, analysis and evaluation involving human subjects or sensitive secondary data. This procedure outlines explicit guidelines for data collection which includes reflection on issues pertaining to data privacy, the rights of children to be consulted on issues which affect them, informed consent, security and confidentiality.

However, with increasing collection of big data and a vocal data science movement calling for more open data and greater utilization of big data within public, private and not for profit policymaking arenas, ensuring the protection of and respect for children rights is becoming increasingly challenging.

With respect to defining 'Big Data', multiple definitions and little consensus exist, The United Nations Global Pulse (2013) highlights the nature and qualities of big data noting that:

Big Data is an umbrella term referring to the large amounts of digital data continually generated by the global population. It refers to the speed and frequency by which data is produced and collected – by an increasing number of sources. [It] generally shares some or all of the following features:

1. Digitally generated
2. Passively produced
3. Automatically collected
4. Geographically or temporally trackable
5. Continuously analysable. (p.3)

While recognizing these characteristics, in this paper we refer to Canavillas (2016) definition of Big Data. This reflects a position that Big Data is a technological phenomenon, in so far as it can be described as:

---

[1] This paper is a reproduction of the original paper published as Berman, Gabrielle; Albright, Kerry (2017) Children and the Data Cycle: Rights and Ethics in a Big Data World, no. 2017-05, UNICEF Office of Research - Innocenti, Florence, available at; https://www.unicef-irc.org/publications/907/

> *Data sets that are so large or complex that traditional data processing applications are inadequate to deal with them. The term 'big data' often refers simply to the use of predictive analytics, user behavior analytics, or certain other advanced data analytics methods that attempt to extract value from data,* (Canavillas et al., 2016).

It should also be noted that, while adopting this definition, this paper also recognizes that big data is not solely a technological phenomenon; it also has cultural and social dimensions relating to expectations of its applicability, robustness, accuracy and objectivity, across multiple domains – ranging from education to justice systems (Boyd and Crawford, 2002). Within this framing of big data, we can truly start to unpack not only the nature of big data and its function but also its implications and potential impacts.

With this perspective in mind, fundamental questions need to be raised as to how best translate universal principles regarding the rights of the child and traditional ethical frameworks for offline data collection, analysis and regulation into an online environment. This includes ascertaining how to uphold such rights, and balancing the risks and opportunities for children that engagement may bring, especially in a world where an estimated one-in-three of all global internet users today is below the age of sixteen. (Livingstone et al., 2015). As noted by Floridi and Taddeo (2016):

> *We have come to understand that it is not a specific technology (computers, tablets, mobile phones, online platforms, cloud computing…), but what any digital technology manipulates that represents the correct focus of our ethical strategies…It is not the hardware that causes ethical problems, it is what the hardware does with the software and the data that represents the source of our new difficulties* (Floridi and Taddeo, 2016, p.3).

Data collection, analysis and regulation in the digital age raises questions about both the realization and the protection of children's rights. It raises questions of whether traditional ethical frameworks that guide academic research in institutional settings – and national legislative frameworks that pertain to data collection and consent from children, are adequate and sufficient. In the first instance, analysis of big data frequently does not occur within the confines of research institutions; it is consequently not bound by human subject protections. Furthermore, big data is frequently collected by both public and private organizations, and is therefore subject to multiple and varying international and state-based interventions and standards. Frequently, there is insufficient guidance, or practical and effective solutions, to safely collect data directly or indirectly from children within a digital world.

The response therefore needs to be both generic and specific: Ethical frameworks for big data collection, retention and analytics, which go beyond traditional research paradigms, are urgently required for the general population more broadly, but also for the child population, specifically. These frameworks are needed to guide institutional, national and international practices throughout the entire data cycle – from collection through to destruction or removal – wherever possible and appropriate.

Finally, in the absence of a narrow linear relationship between data providers, collectors, analysts and users, multiple approaches are required to ensure ethical practices and outcomes. Varying or multiple solutions can, and should be considered, at each stage of the data chain. Solutions to ensure the protection and participation of children will need to explicitly recognize and respond to the reality that research and data collection is no longer bound by the established protocols and operating procedures of the academic community; analysis may be undertaken by people who may not be child rights experts nor trained researchers, familiar with the concept of ethical standards, and may not be bound by notions of the best interest of the child. This may bring the benefit of fresh perspectives, but also significant ethical challenges.

## 2. WHY IS EXPLICIT CONSIDERATION OF THE CHILD WITHIN DATA SCIENCE IMPORTANT? WHAT MAKES CHILDREN DIFFERENT AND DISTINCTIVE?

There are a number of defining features of children and their lives that work interactively to imply that data science needs to explicitly consider its ethical implications for children. The most obvious of these is the growing demand for and use of big data and the rapid development of technologies for its collection and analysis. This accumulation implies that more data will be collected on children over their lifetime than ever before. The result is that the future use, applications and consequent impacts on their lives, is still largely unpredictable.

In short, without broader and coherent ethical frameworks for data science governance children are likely to suffer the consequences hardest and longest. Conversely however, they are also likely to reap the greatest potential benefits. This said, the net impact on children will be determined by our capacity to negotiate this tension and to explore, understand and address potential risks and benefits for this segment of the population.

Traditional ethical frameworks for research and data collection are unquestionably problematic for children growing up within this digital age (Livingstone et al., 2015).This is particularly true in light of the persistence of data collected throughout the probable life course of the child, and the consequent uncertainty regarding the impacts of self-rendered and externally imposed digital identities on the life-long consequences and life choices of children. This uncertainty renders any assessment of potential harm and benefits in the 'best interest of the child' as required by Article 3 of the CRC difficult – if not impossible – in the face of:

(a) Unknown future applications of data (Fossheim and Ingierd, 2015);
(b) Children's and parents' understanding of the implications and applications of their data with the attendant implications for self-management of their digital identities, (Blackwell and Gardiner, 2016); and
(c) The insufficiency of traditional informed consent and assent processes, given the nature of the data collected from the Internet, as well as the frequent opacity of the ages of data providers.

This opacity has implications in so far as it may confound the adoption of a more nuanced definition of childhood in line with the 'evolving capacities' of the child, identified in Article 5 of the CRC and the recently adopted Committee on the Rights of the Child

General Comment No. 20 on the Implementation of the Rights of the Child in Adolescence. Approaches adopted to ensure the realization of the rights of adolescents should differ from those adopted for younger children; recognizing children's development and their increasing competencies, analytical capacities and agency and the implications of consent.

In addition to child-specific rights issues, children may be exposed to many, if not all of the same issues that are present for adults – including the questionable capacity to truly anonymize data and the potential for de-anonymization. This inability to guarantee anonymity may confound the State's obligations to ensure the right to privacy as reflected in Article 16 of the CRC and Article 17 of the International Convention on Civil and Political rights (UN General Assembly, 1966).

Big data may also potentially silence the voice of the child by encouraging the use of data and analytics rather than dialogue and engagement to ascertain perspectives, preferences, attitudes and competencies (Lupton and Williamson, 2017), in direct contravention of Article 12 of the CRC.

Finally and importantly, big data raises ethical issues relating to the increasingly separate and distinct processes and actors involved in the creation and collation of data sets, analysis and use; their varying degrees of knowledge and technical expertise; their divergent interests; and the frequent absence of peer reviews or audits of data and algorithms to determine the validity of both the data and the consequent findings – used to inform decision making. Each of these issues are explored in turn in this paper.

## 2.1 DIGITAL IDENTITIES AND IMPACTS OVER THE LIFE COURSE

One of the most critical issues as relates to Big Data and children is the impact on their digital identities over their life course. As noted by Papacharissi (2010), '[the] networked self is an amalgam of identities that are created across multiple online platforms, constituted via an array of social media tools'. Helmond (2010) adds two concepts to this idea: first, that this identity online is in perpetual beta, implying that the nature of these software platforms results in the acquisition of information (updates, photographs, additional information) ad infinitum, leading to a constantly evolving representation of self. Second that an individual's/child's material online is often generated by other users and the written and visual images provided may have greater impact on an individual's/child's networked self than that which they provided themselves. Hence despite even the most careful 'curation' of one's networked self, the networks can – and do – hold significant power over these identities. A third player in the construction of digital identity is the host of the social media services that utilizes the data, often for economic purposes. Within this context, the data that is collected from children may at any uncertain point in the future be utilized and analysed by indeterminate algorithms, for indeterminate clients, to create digital identities of which the individuals/children are unaware.

This formation of digital identities by corporate third parties can be extended to include not only social media services, but also multiple digital service providers (including government and private parties), who collect, share and/or sell private data. These organizations can retain a range of data including self-tracking data, data collected from the Internet of things, administrative data and data required of children to access targeted child-friendly programmes. As the number of collectors of children's data grows, so too does the possibility of the child and/or their parent losing control over their digital identity. The implications of lack of control over digital identities will be explored in further detail later in this paper. However, the repurposing of data and the algorithms applied, noted in the next section, can have significant impacts. This includes implications and impacts on reputation, access and costs of services, education, employment opportunities and personal security (Pasquale, 2016) to name but a few of the areas where digital identities can and may affect life choices and importantly, the opportunities available to children. While the limits to management of digital identity holds true for adults as well as for children, the impacts on a child's development are less certain given the biological and cognitive changes that occur during the period to early adulthood and the resultant formation of self-esteem, individuality and independence (Eccles, 1999).

## 2.2 CHILDREN'S UNDERSTANDING

Related to the issue of control over public identity formation, is the valid concern that children may not have full knowledge or understanding of the implications of data accessibility and subsequent uses. The notion that children are media-savvy, informed consumers, with a clear understanding of processes, issues and implications, is highly contested (Livingstone et al., 2015, Chung and Grimes, 2005, Valkenberg and Cantor, 2002, Shade et al., 2004). While this generation of children may be adept at utilizing internet technologies, the presumption that the greater proportion have an extensive and nuanced comprehension of issues such as persistence, third party sale of data, analytics and applications, let alone the legal jargon related to data collection for various sites, or the implications of advanced website/browser tracking programmes such as canvas fingerprinting and evercookies (Acar et al., 2014), is overly optimistic. This perspective also fails to account for the ongoing development of children's brains and greater emphasis on shorter-term outcomes in decision making (Reyna and Farley, 2006). Furthermore, while children (and indeed their parents) may be aware of basic privacy settings, (Acar et al., 2014), '*even sophisticated users face great difficulties*'. *(Ibid).*

## 2.3 INFORMED CONSENT

These limitations and concerns are compounded by the fact that traditional modes of ensuring consent and safeguarding child rights are neither possible nor feasible in an online environment. Under many national and international legislative and regulatory frameworks, guardians or parents are responsible for providing parental consent for the collection of data from children under eighteen or the relevant age of majority. Furthermore, and within the context of *UNICEF's Procedure for Ethical Standards in Research, Evaluations and Data Collection and Analysis*, informed consent or assent (where children are not legally permitted to provide informed consent) should be received from the child, following clear articulation and full disclosure of the planned use of the data collected, communicated using language and methods easily understood by children. Essentially, clear guidance must be provided to children to enable them to withdraw from participation or refuse to provide data at any point in the process (pp.11-12, para. iii and iv). It should be apparent that these Standards were primarily developed with more traditional forms of data collection and analysis in mind, which presuppose a linear relationship between those providing the data and the researchers.

It should also to be apparent that there may be circumstances in which children or adolescents would like to give their informed consent or assent, but are unable to do so, if parental opinion differs. Here again, there is a need to consider the evolving capacities of the child (defined as the process of maturation and learning through which children progressively acquire competencies, understanding and increasing levels of agency to take responsibility and exercise their rights (UN General Assembly, 2016). Rather than making a binary distinction between child and adult, a spectrum approach based on evolving capacities, which balances two key considerations, is required. Firstly, measures must exist to guarantee the rights of adolescents to express views on all matters of concern to them, in accordance with their age and maturity, and to ensure their views are given due weight (acknowledging the significant opportunities for strengthening and expanding their engagement that the online environment provides). Secondly, the rights of adolescents to privacy and protection, including in relation to their parents who may frequently have oversight of what their children access or write online, or who engage in 'sharenting'.

Attempts to address issues related to children's consent in online environments have emerged through regulatory frameworks such as the Children's Online Privacy Protection Act (COPPA) (US Federal Trade Commission, 1998) and more recently, the adoption of the EU General Data Protection Regulation (GDPR) (European Parliament and Council, 2016). COPPA includes the requirement that commercial websites aimed at children under age 13, give parents notice about their data collection activities, obtain verifiable consent from parents prior to collection of data from children, provide parents with access to any information collected from children, and finally give parents the opportunity to discontinue further uses of the data collected. Under these regulations however, the onus remains on providers to determine the form and nature of the mechanisms to ensure privacy and the effective informed consent processes. The GDPR explicitly recognizes that children deserve specific protection of their personal data, and introduces additional rights and safeguards for children. It requires parental consent for the processing of personal data of children under age 16, with the qualification that EU Member States may lower the age requiring parental consent only to age 13. However, as noted by Livingstone and Locatelli (2012):

> *Once youth go online, the challenges of obtaining informed consent, already significant for research with children and youth, are magnified. The old adage that "on the Internet no one knows you're a dog" is still pertinent, since on the Internet no one knows if you are a child*
> (Livingstone and Locatelli, 2012, p.68).

Hence, while regulations exist – or are being established – to ensure consent, it would be hard to argue that this consent is truly informed, certainly not in accordance with UNICEF standards, nor is it likely to be comprehensively effective in light of the difficulties in authentication and the variability in oversight. With respect to informed consent, a 2005 study by Chung and Grimes of child-based websites underlines that there are no clear guidelines on the nature of informed consent required on child-based sites and that the terms and conditions on many of these sites may be framed in legalistic language, which is unlikely to be understood by adults, let alone children. These terms may grant websites unrestricted and exclusive use of private data that include not only preferences but also postcodes, names and email addresses. In these contexts, children and their parents are frequently presented with a binary choice to either accept the complex set of terms or to forsake the service in its entirety, since private corporations are entitled to restrict access to sites should parents or guardians of children not consent to complete, unfettered use of the data. Furthermore, as noted by Boyd and Marwick in 2011, data can – and frequently is – used outside the contexts in which they were supplied; in many instances it is entirely possible that neither parent not child would have agreed to such a range of uses. Finally, it is apparent that the concept of 'evolving capacities' of the child in relation to informed consent, needs much greater unpacking and nuancing in existing legislative and regulatory frameworks.

## 2.4 THE PERSISTENCE OF DATA

As noted by Boyd in 2008, the data collected from the Internet (and indeed from other technologies) is characterized by its persistence; in that it is automatically registered and stored. The persistence of data collected presents ethical challenges for both adults and children. However, the enduring nature of this data will impact over more of the lifetime of the children, with significant implications for their public/digital identity, their capacity to shape this sphere, and the longer term impacts and outcomes (Ess, 2015).

With the passage of the GDPR, steps have been taken to allow for the removal of personal information from the Internet. This right is referred to as the right to erasure (Article 17 of the GDPR), also known as 'the right to be forgotten'. This right is designed to enable an individual to request the deletion or removal of personal data whether there is no compelling reason for its continued processing. The right to erasure does not provide an absolute 'right to be forgotten'. Individuals have a right to have personal data erased and to prevent processing in specific circumstances:

- When the personal data is no longer necessary in relation to the purpose for which it was originally collected/processed;
- When the individual withdraws consent;
- When the individual objects to the processing and there is no overriding legitimate interest for continuing the processing;
- When the personal data was unlawfully processed (i.e. is in breach of the GDPR);
- When the personal data has to be erased in order to comply with a legal obligation; and
- When the personal data is processed in relation to the offer of information society services to a child (Information Commissioners Office, 2016).

More specifically, the GDPR introduces extra requirements when the request for erasure relates to children's personal data. Data collectors are required to pay special attention to existing situations where a child has given consent to processing and later request erasure of the data – especially from social networking sites and Internet forums – regardless of age at the time of the request. This is because a child may not have been fully aware of the risks involved in the processing at the time of consent (Recital 65, referenced in Information Commissioners Office, 2016). However, the most compelling argument for successful right to erasure remains the fact the data was unlawfully processed in the first place, due to a lack of free and informed consent.

The GDPR also requires that if the contested personal data has been disclosed to third parties, they must be informed, and are required to erase all links to, copies or replication of the personal data in question, unless it is impossible or involves disproportionate effort to do so. Even though the GDPR was only formally adopted in April 2016 and will not enter into force until May 2018, this requirement has already been pre-tested in a ruling on by the European Court of Justice (ECJ) on 14 May 2014, whereby a Spanish man was able to secure the deletion of information dating back to 1998 by Google Spain, as a subsidiary of Google Inc. (Gibbs, 2015a). This landmark decision by the ECJ is a significant step in ensuring individual control of personal data. This is particularly notable given that Google Inc. who controlled the data, was viewed as an 'establishment' within the meaning of the directive' implying that under the regulations, parent companies must adhere to the provisions of the regulation if their subsidiary is in Europe, despite being located elsewhere (Court of Justice of the European Union, 2014). The GDPR will also refine the limitations to the 'right to be forgotten' established by the ECJ (European Commission, 2014).

Such regulations offer some hope for the control of personal data and particularly – given the specific provisions for children – for the control of children's data, though there are some caveats:. Firstly and importantly, children and their parents need to be aware that this data exists. The frequent lack of transparency around data transmission to third parties, may make it difficult, if not impossible, to determine the nature of the data shared. Secondly, there is a need to establish with whom the data has been shared with and in what form; third parties may in fact sell on raw or processed data and may not even be aware of the multiple agents who have access to this data. Thirdly, the uses of the data may be unknown. Hence while the original data may not be perceived as problematic, or as having the potential to negatively impact on a child's digital identity and privacy, the further analysis of this data (notably outside the context for which it was generated or combined with additional datasets) may do just that. Finally, it should be emphasized that regulations such as the GDPR are localized and not universal. Significant further work is thus required to ensure that these positive protective measures are more globally applied. These issues will be discussed and elaborated in the proceeding sections.

## 2.5 DATA ANONYMIZATION

One of the more concerning aspects of existing regulatory frameworks designed to ensure children's privacy and provide general protections is that they frequently fail to require minimum standards (beyond loose prescriptions of age of consent and post hoc requirements for removal of data). This is particularly problematic, given the speed at which technologies and solutions develop. This is compounded by the fact that the capacity to ensure privacy through anonymization and aggregation is highly questionable (Steen-Johnsen and Enjolras, 2015; Boyd and Crawford, 2012) – although experts are divided about the level of risk in practice. The proliferation of technologies that provide geographical positioning, metadata such as email addresses and the increase in the linking of databases as a result of the integration of social media and other internet sites such as Google, Gmail, YouTube, Chrome and Google+, allows for the potential creation of very detailed information on individuals. While data may initially be sold on or provided to a third party in aggregate form, this does not preclude the potential for de-anonymization or disaggregation of data, (Boyd, 2008) or the so-called 'mosaic effect' (Howard, 2013), particularly if clauses on websites allow for unrestricted future uses.

This is not to deny that enhanced open and linked data, particularly government data, can lead to many potential benefits for citizens – including children – such as improved access to health care and better delivery of public services (Open Data Institute, n.p.). However, while much government-collected data contains personally identifiable information (PII), governments are generally obliged by privacy laws to avoid disclosing personal information, except for authorized purposes that could allow for its use in restricted or de-identified forms. Generally speaking, for human subject data to be "open", it needs to be based on informed consent from the participant, which for most open data purposes – including satisfying IRBs and ethics boards, is not a case of opting-out but of opting in. This is clearly often not the case in most big data environments, where much data is passively collected and choices to sell on data are several steps removed from the original collection of data.

As already mentioned, there is no broad consensus on the potential risk of the mosaic effect, or on the potential and limits of de-identification technology (Shaw and Cloud, 2014). However, the term 'anonymized data' is often used to imply that the data can no longer be re-identified. However, most experts agree that data anonymization is not foolproof and that there is a tension between utility and anonymity: Data can often be either useful, or anonymous, rarely both. Techniques such as 'differential privacy' can go some way to ensure privacy protection and prevention of misuse of data (Center for Open Data Enterprise, 2016), but much greater regulation is needed in this area.

This capacity for data to be re-identifiable has the potential to impact children throughout their life cycle in negative ways. This lack of control of their public identities could potentially impact their access to educational, employment and financial opportunities, enhance their potential exposure to discrimination, and at the more extreme end of the spectrum, allow political actors to use this data to assert control over their lives and regulate their personal and political expression. While these negative experiences are not limited to children, this generation will be the first to experience these issues throughout their life cycle, and particularly at early life stages and critical junctions in their personal development and public life. Furthermore, ensuring privacy – even with appropriate anonymization in longitudinal data – is also extremely difficult, given the fact that such data will have multiple transactions per individual; hence indirect identifiers will be greater than eight – the recommended maximum to prevent re-identification (El Emam, 2016). As technology develops and more information is captured, commodified, analysed and applied, there are likely to be greater possibilities for this data to be misused and for this generation and the generations that follow to be exposed to higher levels of risk and greater violations of basic rights. A recent paper from the National Institute of Standards and Technology provides a thorough review of types and limits of de-identification (Garfinkel, 2015). However, given the current level of uncertainty around being unable to ensure continued data privacy and the particular vulnerabilities of children, we need to err on the side of caution with data generated on and by children, at this stage.

## 2.6 UNKNOWN FUTURE APPLICATIONS AND USE

These concerns regarding children's present and future rights and their capacity to control their digital identity, is complicated by unknown technological developments and the opacity of current and potential applications and uses of contemporary systems. One of the most common directives relating to ethical oversight of data collection is the need to maximize benefits and to minimize risks to participants with a minimum standard of 'do no harm'. International and national ethical guidelines such as those produced by CIOMS and WHO (CIOMS and WHO, 2002) and those from the United States (United States' National Commission for the Protection of Human Subjects of Biomedical and Behavioral Research, 1978 and the US Homeland Security, Science and Technology Division, 2012) note the need for more stringent decision-making algorithms on whether or how to collect data from vulnerable populations, including children. This requirement is also clearly articulated in existing guidelines that explicitly pertain to data collection and research involving children (UNICEF, 2015; Graham et al., 2013).

However, as already noted, big data collection can present significant hurdles in assessing future potential harm and benefits to children; not least because the nature of much big data is passive data collection that does not allow for explicit consideration of these issues at collection stage. Even after initial collection, there are major challenges to assessing potential future harm and benefits due to uncertainty in technological advances, and to current arrangements that limit the control of personal data, by allowing organizations to retain long-term rights over data and its use. Furthermore, international ethical guidelines such as CIOMS and WHO (2002) frequently require community consultation to understand the impacts of dissemination of datasets on children and their communities, and to ensure appropriate representation and use. These remain relevant and critical, but are nearly impossible to enforce in a big data world.

## 2.7 SILENCING CHILDREN'S VOICES

A further concern applicable to adults but with particular salience for children, is the potential for decision makers to substitute direct dialogue and engagement with children with the cheaper and quicker approach of passive, big data collection. The replacement of engagement with algorithms, as noted previously, is in direct contradiction to Article 12 of the CRC, which clearly articulates children's rights to have a say in matters that affect them. This is reinforced in the General Comment No. 20, which articulates that in line with the 'evolving capacities' approach, adolescents in particular have a right to take increasing responsibility for decisions that affect their lives (UNGA, 2016). While arguments can be made for the value of algorithms in burdened child protection, justice and educational systems, serious concerns remain regarding the implications of the potential omission of children's voices in these domains. As noted by Lupton and Williamson (2017):

> *The embodied and subjective voices of children [may be] displaced by the supposed impartial objectivity provided by the technological mouthpieces of data.*
> (Lupton and Williamson, 2017, p.11),

Beyond Article 12, a further implication of silencing children is the 'supposed impartiality' of data and its ability to accurately assess competencies, preferences, future actions and to correctly assign children to various categories such as 'at risk' or 'likely to re-offend'. Even assuming high degrees of predictive accuracy. Any truly ethical and moral framework would need to consider the implications of errors at the margins. This is requisite if we are to pursue not only the best interests of most children, but also the best interests of the 'child' overall, which includes the 'outliers'.

## 2.8 FRAGMENTED SYSTEMS OF DATA COLLECTION, ANALYSIS AND USE AND LACK OF PEER REVIEW AND AUDITS

The potential impacts of big data are further problematized by the current reality of fragmented ownership of data and heterogeneous regulatory frameworks. This fragmentation of ownership across both the public and private sectors, as well as across geographic spaces, implies that decision-making regarding collation and creation of data sets, analysis and use, is frequently not reviewed in a public forum through stakeholder consultation, peer review or a formal ethical review process, as is usually mandatory in conventional offline research. This is further exacerbated by the fact that even in contexts such as academia, where formal oversight and ethical reviews take place, criticism has emerged regarding participant expertise and the appropriateness of the standards and processes adopted to analyse research programmes involving technologies that utilize big data (Future of Privacy Forum, 2015).

This situation provides challenges in terms of control, access and assessment of the quality of the data and its conclusions. The margin for error, abuse and misapplication and how this impacts children's lives is unknown but alarming. Furthermore, the capacity of existing regulatory frameworks to mitigate potential risks is problematized by the separation of actors within the data chain, as noted by Prabhu (2015):

> *Data collection, curation and analysis do not necessarily take place at a single point which can be subjected to robust regulatory measures. Moreover, the technical opacity of algorithms underpinning Big Data analysis, as well as the real-time nature of such analyses, does not easily lend itself to meaningful scrutiny by way of traditional transparency and oversight mechanisms.*
> (Prabhu, 2015, p.166),

Manovich, as cited by Boyd and Crawford (2012), identifies three key players within the sphere of Big Data: the creators of data; the collectors of data; and the experts who analyse the data. According to Boyd and Crawford (2012), the experts who analyse the date are the smallest and most privileged group, in so far as they determine how Big Data will be used and who gets to participate. It may be argued however, that the final end-users of the data are excluded from this list of players. They are the ones who will determine the ultimate application of the data, if not the determination of the algorithm itself. The inclusion of this additional player may be critical to determine appropriate means to ensure ethical data collection, analysis and use and to facilitate outcomes that support – rather than detract from children's rights – in the use of big data. We will return to this issue in subsequent sections but first it is important to understand how the growing collection, analysis and use of big data in contemporary society is impacting the preservation of or respect for children's rights.

## 3. BIG DATA AND CHILDREN: THE GOOD, THE BAD AND THE UNKNOWN

Taking the 1989 Convention on the Rights of the Child as a starting point, it is important to recognize the particular vulnerabilities of children and the special protections they should be afforded, the significance and importance of data analytics in contemporary society, and the degree to which this discipline is intimately entwined with ensuring the preservation of and respect for children's rights.

### 3.1. THE GOOD

Undoubtedly, developments in the platforms for the collection and subsequent analysis of big data have some clear and obvious benefits to children, relating to their protection, their safety and their participation within the broader global community (Livingstone et.al., 2015). Such benefits include the creation of platforms that enhance children's rights with respect to access to information (Article 17 of the Convention on the Rights of the Child) and the provision of vehicles to facilitate freedom of expression (Article 13). Furthermore, and importantly, the platforms and software used to collect, collate and analyse big data have many other potential direct and indirect benefits for children, in areas as diverse as child survival and development, improved access to services, the prevention of violence, and early warning detection of natural and other hazards (UN Global Pulse, 2013).

Such technologies include crisis mapping platforms, which gather crowd-sourced data from mobile phone users, in humanitarian contexts. These technologies and the data produced facilitate the mapping of incidences of violence or disasters, enabling appropriate responses, support and the dissemination of information to affected parties. A well-known example is the Ushahidi platform, which was first designed to map reports of post-election violence in Kenya, in 2008, and was subsequently utilized in the post-earthquake response in Haiti (Moestue and Muggah, 2014). The platform enables crises mapping and identification of areas of urgent humanitarian need, as well as providing real-time information on locations of violence. It has the potential to be applied to assess patterns and trends of violence against children and to inform possible solutions targeting volatile locations (Moestue and Muggah, 2014).

Another example is the collection of data by Unmanned Aerial Vehicles which can provide real-time information and situation monitoring, public information and advocacy, search and rescue, and mapping. The OrUAV developed by Google and aid agencies has been used to identify locations to drop deliveries in emergencies to address the very real needs of children, particularly those under 5, who are usually the first to become malnourished and die (Moestue and Muggah, 2014). This remote provision is particularly valuable since it not only ensures targeted provision of life-saving goods, but also has the potential to prevent loss of life amongst those aiming to respond to the initial emergency. At the global level, big data is also being collected and analysed as part of the Google Global Human Trafficking Hotline Network. This links local, regional and national anti-trafficking helplines, collecting data across the network. Analysis of helpline datasets allows for the logging of incidences of victimization and the mapping of distribution of aid resources and services (Moestue and Muggah, 2014).

Data analytics have also helped mitigate some of the most insidious side effects of Internet communication on children. Technologies developed by Microsoft and Dartmouth College are used to 'tag' child abuse images both online and in cloud based storage, thereby allowing law enforcement and other agencies to rapidly identify and detect any reproduction of these images – even if they have been slightly altered, preventing them from being uploaded again or expediting their removal and investigation if they have been stored (Ith, 2016). In a similar vein, the FBI has developed and uses software called the '"Network Investigative Tool" to collect identifying information of those accessing servers that distribute child abuse images uploaded on the 'dark net'.

These examples highlight just a few of the ways in which this confluence of technologies and analytics may positively impact on children's protection, their development, wellbeing, participation, inclusion and access to services. What should be evident however, is that within each of the above contexts, big data was or is being used with a clear and explicit focus on supporting and protecting individuals, communities and their children. These potential benefits must be considered alongside the potential use of big data for purposes other than the health and wellbeing of data providers and their communities; for example, in the context of the potential for repurposing of data for less humanitarian ends, and, in circumstances where uncritical application of algorithms and use of technologies to collect this data may result in unrepresentative or inequitable outcomes.

### 3.2. THE BAD

The capture and use of big data, however, also raises significant concerns relating not only to privacy and loss of control of personal data, but also to the potential for direct or inadvertent discrimination and profiling, scope creep and technological dependency – resulting in restrictions on access to vital services.

Children are increasingly contributing to online content, through online environments, games and discussions. (Chung and Grimes, 2005) While the participatory aspect of online environments can and should be lauded as a tool for supporting access to geographically disparate communities, information, recreation and educational opportunities (Livingstone et. al., 2015)., the value the online environment affords must also be considered alongside the counter-opportunities for private and public organizations to collect big data on children and the frequently unknown, subsequent uses of this data.

The collection of children's data by a broad range of actors presents legitimate ethical issues regarding the capability of organizations to maintain the privacy of individual children as required under

Article 16 of the Convention on the Rights of the Child. According to Ghosh (2015) a flaw in a children's online website left 3.3 million children's personal details, registered with the site, vulnerable to hacking. In a further case described by Ghosh (2015), children's data was stolen from a toy maker's website. This resulted in the capture of the private data of 6.4 million children including their photos and physical addresses.

The collection of data on children is further problematized by the on-sale and sharing of data with third parties, primarily for marketing purposes, but also for alternate uses – known and unknown, none of which are necessarily driven by a directive of the best interests of the child, nor are necessarily open to scrutiny in the public domain. Perhaps the best-known example is that of the controversial 'Smart Barbie' doll produced by Mattell, which led privacy campaigners in 2015 to highlight that recordings of children using voice recognition technology were being sent to third-party companies for processing, potentially revealing his or her intimate thoughts and details (Gibbs, 2015b). All of these data collection methods have the potential to limit the control children have over their information and their public identities. Data mining technologies can create detailed demographic and behavioural profiles of children online, raising issues of privacy and intellectual ownership.

The potential for data mining to give rise to discrimination is another concern. A literature review regarding the potential for discrimination, arising from big data mining by Barocas (2014), identified three means by which discrimination may occur. Firstly, conscious discrimination may occur, which may be difficult to discern by virtue of the use of algorithms that are premised on underlying factors that may define a particularly vulnerable cohort, such as geographical location or health profile. Secondly, discrimination may result from proportional misrepresentation (under or over representation) of marginalized groups within a particular sample, leading to inaccurate conclusions, rankings and skewed decision making. Finally, discrimination may result from over dependence on specific data sources for decision making - to the exclusion of more verifiable, or nuanced approaches, or the utilization of multiple methods, to allow for triangulation of the data.

According to Nissenbaum (2009) the significant driver of discrimination occurs when data is moved out of context, and the contextual integrity of the data is compromised. Pasquale (2014) notes the proliferation of poorly regulated data miners, brokers and resellers, who are providing varied categorizations of persons on a breadth of issues ranging from HIV status, to mental health status, to exposure to sexual abuse. He highlights that these categorical lists raise three ethical issues: such lists are frequently inaccurate and almost impossible to verify; they can – and are – inappropriately used for decision making; and people are most likely to be unaware they are on these lists.

Encompassed within this type of discrimination, is the use of big data for predictive analyses ('predictive analytics'), particularly as it pertains to the identification of 'at risk' youth. Techniques such as predictive risk modelling (PRM) use huge volumes of historical data to evaluate the likelihood of negative events in the future. Using PRM, social service agencies are able to crunch through vast amounts of old case data to provide predictions about which children may face the greatest risk of future harm. The approach – already in widespread use in health care and policing – holds tremendous appeal, especially for cash-strapped social service agencies; it can flag the highest-risk cases for intervention by always-too-few case workers. Initial pilots in countries including New Zealand and the US, have, however, raised concerns about this approach, in terms of child protection. Issues of data privacy, the underlying drivers of abuse and neglect, and systemic biases, have all been raised by the concerned groups, which include UNICEF New Zealand (Le Goulven, 2017).

Pasquale (2015) notes the use by academics of poorly regulated scoring services, to identify potential 'problem students' based on calculations the students cannot access and of which they are unaware. Tested assessment tools are currently being used or explored in the juvenile justice system, to determine the likely recidivism of juvenile offenders (Judicial Council of California, 2011). However, even the most publically available and validated tools, are providing mixed results. As noted by the Judicial Council of California (2011), given the mixed findings from the validation studies on these instruments, and the limited research currently available, the results from these tools should not be used as the sole determinant of a young person's risk of sexual re-offense (p.4). The use of predictive data in the juvenile justice system is a cause for concern. While the Judicial Council of California (2011) highlighted the need for a cautious and qualified use of tested tools, the reliance on big data and algorithms to determine 'at risk youth' have very significant implications for the treatment and sentencing of young people.

It should be noted that many of these issues would be picked up in technical and ethical reviews of traditional research – most notably the potential for sampling bias, the appropriateness of the analytics, and issues relating to the robustness of the datasets. However, because of the frequent opacity in many private and public institutions regarding the use and nature of the algorithms that are applied and the databases that are mined, this form of discrimination is frequently impossible to assess. While these concerns apply equally to the use of online data generated both by children and adults, it is argued that the additional duty of care and protection afforded to children in traditional research, remain valid. This interpretation is not only in keeping with legal and institutional frameworks for the protection of children, buts also clearly reflects the potentially longer-term impacts of discrimination on children's opportunities and life choices.

### 3.3. THE UNKNOWN

From the preceding sections, it should be evident that the integrity of findings from the analysis of big data used for decision making may not be assured, due to the possibility of manipulated, un-critiqued or opaque algorithms; biased interpretations; and poor quality or unrepresentative data. While this can also be said of traditional data collection systems, the peer review system and established systems of ethical standards go some way to mitigating this. However, in the era of big data, the three core strategies long-used to ensure privacy: individual notice and consent; opting out; and anonymization, have lost much of their effectiveness. (Mayer-Schönberger and Cukier, 2013, p.156)

Furthermore, the value of information no longer resides solely in its primary purpose, but also in potential secondary uses or 'interoperability' of data. In a big data age, even if the notion of informed consent is possible, when the data are first collected, their most innovative secondary uses cannot be imagined. How can organizations provide notice for a purpose that does not yet exist?

How can individuals give informed consent to an unknown? In the context of big data, the tried and trusted concept of notice and consent is often either too restrictive to unearth the data's latent value, or too empty to protect an individual's privacy (Nissenbaum, 2013, p.154).

A lack of knowledge about the future purposes and uses of data is particularly concerning in socially and politically volatile circumstances, with technologies and data is susceptible to misuse or misappropriation by repressive State actors or authoritarian elements (Hosein and Nyst, 2013). Whilst big data has a huge potential to be used for social good, proactively preventing new modes of discrimination that some uses of big data may enable - particularly with regard to civil and human rights protections - is critical.

In a big data world, children are as susceptible, if not more susceptible than adults, to the long-term ramifications and inappropriate applications of data, while safeguards, security systems and regulatory frameworks attempt to catch up with the technologies and applications, and while users of data are gradually educated on the flaws and potential biases inherent in particular algorithms ("algorithmic discrimination").

The persistence of data and its unknown future applications highlight the limitations of particular ethical and regulatory frameworks in the protection of children's data. The issue remains of how best to ensure that data is processed and utilized in a manner that is consistent with the best interests of the child. This is especially true when child data providers and parents are frequently unable to access and control their data and where the providers, collectors, analysts and users are not in regular dialogue and have varying degrees of access, understanding, technical knowledge and agendas. Furthermore, ethical challenges also exist in ensuring that data and information reduce inequalities, in an era of growing information asymmetry or 'digital divide', in which specific cohorts of children may lack access to appropriate technologies, or may be invisible in datasets and cut off from the potential benefits of the 'data revolution'.

Ensuring children's rights are realized and protected within a big data world requires multiple responses from various actors at each stage of the data chain, and clear and concerted efforts to understand the particular needs and protection that should be afforded to children.

## 4.WHERE TO FROM HERE? HOW DO WE MOVE FORWARD?

> *We cannot have a system, or even the appearance of a system, where surveillance is secret, or where decisions are made about individuals by a Kafkaesque system of opaque and unreviewable decision-makers. (*Reyna and Farley, 2006, p.43)

The big data world requires an explicit focus on child rights and data science, both as a separate discourse and as part of broader discussions on ethical and legal frameworks for big data collection, analysis and use. This discourse needs to take place within a system of multiple actors, including data producers (children and parents), collectors, analysts, end-users and child rights advocates, reflecting on multiple approaches to support both individual agency and societal accountability. Within this system of multiple actors, new forms of accountability and concepts of privacy and consent are required – together with better education for all stakeholders, better regulatory systems that specifically address concerns related to children's data, and better dialogue between stakeholders. The following section therefore provides some considerations on possible mechanisms to support children's rights at all stages of the data chain by the various stakeholders.

Diagram 1 provides a very basic framework of typical players in the child data cycle. It situates these players within the broader ecology of institutional and government regulatory frameworks that have the potential to impact the actions of each player in the cycle. It should be noted that the list of players in each point in the data cycle is not exhaustive, but indicative, and that the players may not be confined to a single role in the cycle but may, in fact, play multiple roles (e.g. the Government as both a collector and user of data and potentially, as a data regulator).

**Diagram 1: Typical Players in the Child Data Cycle**

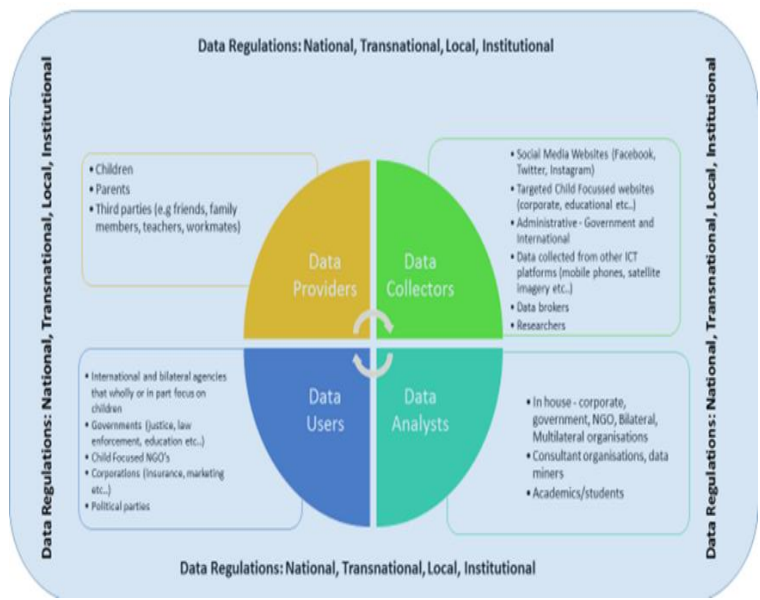

### 4.1.DATA PROVISION – CHILDREN, THEIR PARENTS AND GUARDIANS AND OTHER RELEVANT THIRD PARTIES

The primary stakeholder to consider is the child or their parent or guardian. The main ethical concern for these stakeholders is privacy and control. Targeted interventions to empower this group include both education and consultation. Life-long learning models on ICT, cyber safety and privacy in a big data age should be routinely integrated into generic school-based ICT programmes. Targeting parents as well as children is critical, since they may be the gatekeepers or the providers of children's data within health, education, social media and other settings (Palfrey and Gasser, 2008). Similarly, relevant third parties such as friends and family members should also be considered providers of data when addressing awareness of the risks and responsibilities involved when processing/providing third-party data without consent.

In practical terms, however, it should be acknowledged that such measures are limited, since awareness of privacy-enhancing technologies and cyber hygiene (Sadowski, 2013) struggle to keep pace with tracking mechanisms and Privacy-eviscerating technologies as they develop (Pasquale, 2015). Within this context, the education is valuable in encouraging critical thinking and broad awareness of these issues and their complexities. Importantly, education may be seen as a necessary pre-condition to encourage reflexive thinking on privacy, big data and its implications and to foster civic participation and engagement in ethical debates on these topics.

Building on the notion of engendering civil participation, is the need for fora to encourage this dialogue. Children and parents or guardians need to be consulted on the establishment of regulatory frameworks and the design of online child-friendly consent and privacy settings. The recent endorsement of the EU General Data Protection Regulation (GDPR) is a case in point. The decision to ban internet access providers from turning on child protection filters by default in the name of net neutrality was taken without significant consultations with children, parents or child protection experts (Carr, 2015). Initiatives such as the Pan EU Youth's 2015 Youth Manifesto for A Better Internet (Pan EU Youth, 2014) are laudable. However, additional and more specific consultations and manifestos regarding big data and its uses must be undertaken, regularly reviewed and revised, as technology, education and thinking in this domain progresses.

The fact remains that there are very few direct provisions for children in existing regulatory frameworks and data protection directives. Where they exist, such provisions tend to rely exclusively on parental consent, with little distinction made between older adolescents and younger children. As Macenaite argues in her 2017 article that looks at the attempt by the EU GDPR to adapt children's right to privacy to the digital age,

> *There are specific dilemmas that the introduction of the child-tailored online privacy protection regime creates – the 'empowerment versus protection' and the 'individualized versus average child' dilemmas. It concludes that by favouring protection over the empowerment of children, the Regulation risks limiting children in their online opportunities, and by relying on the average child criteria, it fails to consider the evolving capacities and best interests of the child.*

Achieving an appropriate balance between child protection and participation is not straight-forward. Greater consultation with children and their parents in developing more nuanced regulatory frameworks is, nevertheless, an essential first step.

### 4.2. DATA COLLECTION

Data collectors are without doubt the secondary gatekeepers of children's privacy; regulatory frameworks are required to ensure appropriate data stewardship. At this point in the data chain, it is primarily the nature of the data collected and the control that providers have over their data which must be considered the primary entry points for interventions.

More coherent regulatory frameworks, together with relevant, contemporary notions of consent and requirements for greater information on data use, are needed. Standard international and national privacy frameworks – in a similar vein to the GDP – in recognizing the specific vulnerabilities of children, need to explicitly reflect on appropriate interventions or approaches to dealing with children's data, including in online environments.

As highlighted by Ghosh (2015), stricter regimes for ensuring privacy and security on child-based websites are necessary, and serious consideration must be given to regulatory frameworks related to the nature of information that can be requested from children, at least with respect to personal and identifiable information. Clearer, easily-accessible and understandable terms and conditions from those collecting data, could also be required, including child-friendly descriptions that explain the nature and the form of data being sold. Alternatively or additionally, if child rights are to be truly ensured, options and mechanisms for the removal of data from datasets on request must be developed and included by default. The right to data portability as described in Article 20 of the GDPR should also be considered. This would "allow the data subject to have the right to receive the personal data concerning her or him, which she or he has provided to a controller, in a structured, commonly used and machine-readable format, and to have the right to transmit those data to another controller, without hindrance from the controller to which the personal data have been provided". Such transfer of personal information from one provider to another is of particular relevance when discussing young people and social networks.

Furthermore, clear policies and mechanisms regarding the collection and dissemination of children's data and data in general, are required. More particularly, those organizations that collect data from children or are likely to accumulate significant data from children should be required to ensure accountability to the public with appropriate oversight and protection of privacy, possibly with clear and public statements and guidelines as to the types of organizations to which data has been sold. This could include periodic audits by independent, impartial and professional third parties, or mandatory requirements for the utilization of software – such as PhotoDNA – by data collectors.

### 4.3. DATA ANALYSIS

Data analysts, as noted by Boyd and Crawford (2012) are powerful, critical players in ensuring ethical outcomes in the application and use of big data for children. These players determine (either directly or indirectly) the categorization of individuals and the nature and form of data disseminated. The capacity of these stakeholders to understand the implications of their algorithms and to explain the limitations of both the data and the algorithms adopted to users or decision-makers may go a long way towards supporting more ethical outcomes of big data use. The communication skills of these key players and the capacity to explain and discuss these limitations with users in a clear and simple manner, both pre- and post-analysis, is therefore critical. To this end, education by, and conversations with child rights advocates are necessary, particularly when working on data that will likely impact children and their lives. Whether by design or ignorance, failure to account for the implications of algorithms on children's lives has the potential to negatively impact children at early stages in their life cycle, with further impacts in the longer term.

In order to address this, ethics must become an intrinsic component of all undergraduate and postgraduate courses in this field, and not be left to a few lone pioneers. Furthermore, this ethics component must be applied; teaching students to interrogate algorithms and data outcomes, to determine unconscious and conscious assumptions and potential limitations and ethical implications of their use. In this manner, data analysts can become reflective practitioners. In addition, engaging data analysts in debates with users and producers has the potential to facilitate the development of tools and methods to increase privacy. The development and utilization of such methods offers real opportunities to ensure privacy, while maintaining statistical integrity (Prabhu, 2015). Of all the stakeholders in the data cycle, it is the data analysts alone who will be able to develop or adopt mathematical methods to protect privacy or determine solutions to address errors in datasets.

## 4.4. DATA USE

The final stakeholder in the data chain is the data user. Within government departments, multilateral and bilateral agencies, NGO's and corporations, basic data-literacy skills must be enhanced to raise awareness of the value and the limitations of data, platforms and algorithms. This awareness is critical, particularly given that the technical and privacy aspects of both the hardware and software and the technical considerations needed to determine the robustness of data and the analysis are often poorly understood by users.

As part of this movement towards greater awareness, it must be understood that traditional review processes, premised on human subject research, need to be re-conceptualized, so that the use of technologies and 'publically available' data does not obfuscate institutions' obligations to conduct ethical reviews. Institutions grappling with the ethics of big data and children will need to consider moving away from traditional notions of tangible harm, loss, or negative impacts, towards more inclusive definitions that reflect notions of dignity-based theories of privacy harm. This approach would allow organizations to focus on human dignity and the need to create environments that support personal control and flow of information, without exposing the child or individual to obvious and immediate harm.

With this in mind, institutions need to better unpack expectations and understandings of both children and parents, with regard to what is considered 'personal' and what is considered 'private' (Berman, 2016; Hinton, 2013). A body of research has emerged regarding risks, yet these risks are frequently defined by adults and often reflect pre-established norms and adult concerns regarding the nature of these risks. To date, little has been done to unpack the younger generation's concepts of privacy, expectations of use and notions of public information. This situation is compounded by the limited understanding of the array of risks and opportunities of big data.

Data users should not assume that data analysts understand and will address the issues raised by the use of particular algorithms or datasets, and should, where relevant, seek input from experienced peers or advocates who may better understand the implications. From a child rights perspective, child advocacy organizations who are data users, should not be abdicating responsibility for data collection, analysis and oversight to external, independent data analysts, if they are truly interested in children's outcomes. Data users need to be able to interrogate and/or appreciate the nature of the data used, the security of technologies adopted, and the reliability and application of data, including its limitations and potential impacts on children. Furthermore, child advocacy organizations must be included in discussions with government and private data users, regulators and analysts in order for 'the best interests of the child' to be genuinely considered, and reflected upon, throughout the data cycle. .

Child-focused institutions can, and should, also play a dual role of education and research to re-frame notions of privacy, risk and harm. This approach can better ensure that the understanding, positions, concerns, values and priorities of those most likely to be impacted by them over the life course, are reflected in institutional frameworks. Within a context of limited understanding of the specific mechanics and intricate workings of big data and all its privacy implications, these issues need to be presented to young people to allow and encourage their reflections on what is acceptable in terms of privacy, data ownership and sharing. It cannot be presumed that those who were not 'born digital' understand the reality experienced and the consequent attitudes, perceptions of risk and understanding of privacy, held by younger generations. Child informed frameworks that start from the premise that the adults whose responsibility it is to develop these frameworks should not be exclusively and paternalistically defining risks and prescribing responses, are required. If we fail to acknowledge our limitations, our frameworks will be constrained by our pre-conceptions, our research will be limited to reinforcing these potentially anachronistic understandings, and we will fail to truly protect children in ways and means that are responsive to the contemporary environment and their realities.

With respect to internal governance systems, data users should be encouraged - if not required to regularly disclose the nature and use of algorithms applied to children's data. The introduction of regular audits of the data and the algorithms themselves may also go a long way in ensuring transparency, validity and equity (Crawford and Shulz, 2013). In the case of disputes, independent evaluators could be invited to analyse the selection of data sources, the choice of analytical and predictive tools, including algorithms and models and the interpretation of results (Nissenbaum, 2009). Clear, publicly available organizational protocols, procedures and policies that explicitly focus on the use of big data in organizations that are data users (including policies on the removal of individual data from datasets), would also be a necessary step in ensuring greater organizational accountability and transparency, and could go some way in re-establishing control for children and their families over their private data. The establishment and use of Big Data risk assessments tools for all Big Data usage[2], should be included in these processes While the management of all potential risks may be impossible, the use of these types of tools as a minimum requirement is necessary to ensure clear consideration of issues such as the potential for re-identification, the security and need to access sensitive data, the validity and applicability of any previous broad consent provided as a third party user, data quality considerations and relevant data legislation and security.

---

[2] For an example, see Global Pulse and UNDP (2016) *A Guide to Data Innovation for Development: From Idea to Proof of Concept*, accessed at http://www.undp.org/content/undp/en/home/librarypage/development-impact/a-guide-to-data-innovation-for-development---from-idea-to-proof-.html, p.73.

While risk assessment processes, security measures, consultations, transparency, accountability, education and technical mathematical solutions may provide some possible responses to the ethical dilemmas presented by big data, and more particularly for data provided by or impacting on children, they are neither universally applicable nor exhaustive. The authors argue, however, that the ideas that underpin these approaches remain relevant, namely that responses should target the multiple stakeholders in the data chain, that greater dialogue across stakeholders including children, parents, regulators, child advocates, users and data analysts is required, that technological solutions should continue to be sought, and that an integral component for all actors is education.

**Diagram 2: Possible approaches to addressing ethical issues relating to the child data cycle**

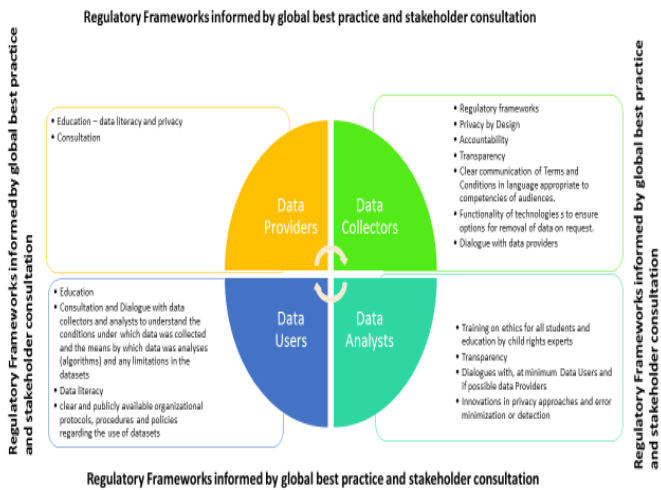

## 5.CONCLUSION

The opportunities presented by big data are considerable. To realize the benefits of big data, the international community must simultaneously address serious concerns about how to protect fundamental rights and values, particularly for the world's most vulnerable populations, including children and adolescents. Traditional, offline ethical standards for research must either be revisited ,or supplemented to reflect data collection from children in online environments, in accordance with the evolving capacities of the child and must acknowledge the implications of the creation and use of 'big data'. Within these frameworks, there is a need to explicitly require increased transparency, accountability, awareness of the risks, the harm, and the benefits associated with big data use. These frameworks must institutionalize the imperative to consider a range of methods to ensure privacy - for instance limiting, wherever possible, the personal data sought from children and encouraging the development and use of privacy enhanced technologies and anonymization techniques. Furthermore, the notion of voluntariness needs to be translated into the digital world, so that children and their families can easily withdraw from ongoing data collection and sharing processes. At the same time, greater reflection is required on the rights of children and adolescents, their right to express themselves and to be heard and importantly their right to privacy and confidentiality - including from their parents as well as from more traditional players in the data cycle. Undoubtedly, consistent international cooperation and guidance in the development of these frameworks and standards is needed to provide clarity on joint adoption and enforcement of applicable rules, standards and best practices (Nissenbaum, 2009).

There are some interesting prototypes that are already being developed in some of these areas, notably on the issue of data privacy. For example, the United Nation's 'Global Pulse' initiative is currently experimenting with development of a tool for assessing the risks, harm, and benefits of Big Data use in global development or humanitarian contexts. This 'Privacy Impact Assessment' approach (UN Global Pulse, 2013) tries to draw attention to whether the data is being used, for example, in a justified balanced and equitable way. . Global Pulse has also drafted some 'Privacy and Data Protection Principles' (UN Global Pulse, 2015) which are currently being debated widely. UN Global Pulse has also established a Data Privacy Advisory Group, which convenes experts from the public and private sectors, academia, and civil society, in a forum to enable continuous dialogue on critical topics related to data protection and privacy, with the objective of unearthing precedents, good practices, and strengthening the overall understanding of how privacy protected analysis of big data can contribute to sustainable development and humanitarian action. Another initiative worthy of mention is the *Policy on the Protection of Personal Data of Persons of Concern to UNHCR,* which outlines basic principles of personal data processing, rights of the data subject, data processing by UNHCR and implementing partners, transfer of personal data to third parties and general accountability and supervision regarding such data (UNHCR, 2015). A UN-wide position paper on Big Data and Privacy is also currently under development.

In 2014, US civil society also made some positive steps with the development of the Civil Rights Principle for the Era of Big Data (The Leadership Conference on Civil and Human Rights, 2014). In Norway, the Norwegian Data Protection Authority, co-sponsored by several other States has called for a 'Big Data Resolution' demanding greater attention to key privacy principles such as purpose limitation, the need to obtain valid consent, the requirement for privacy impact assessments where necessary, privacy by design where feasible and consideration of where anonymization can improve privacy practices (Norwegian Data Protection Authority, 2014). Similarly, a recent UN report on 'The right to privacy in the digital age' notes that the report "may be the first step into realizing an additional protocol to Article 17 of the International Covenant on Civil and Political Rights (ICCPR) to create globally applicable standards for data protection and the protection of privacy in accordance with the rule of law" (UN High Commission for Human Rights, 2014).

However, while international and national frameworks and principles are being developed, issues such as data sovereignty, data quality and integrity and nuanced and robust legal and regulatory frameworks remain an ongoing challenge for all governments and citizens alike. On all of these issues, the voices of children and child rights advocates should be at the centre of these debates; yet these are currently, woefully under-represented. Having celebrated the 25th anniversary of the Convention on the Rights of the Child in 2015 and in light of the UN Secretary-General's call for a 'data revolution' to enhance delivery of the new Sustainable Development Goals (UN Department of Public Information, 2014), there is no better time to encourage greater debate and dialogue between the child rights and data science communities for the betterment of the lives of children worldwide, than now.

## 6.ACKNOWLEDGEMENTS

The authors would like to thank Mario Viola de Azevedo Cunha from the European University Institute and Toby Wicks from UNICEF, for their useful insights and review of this paper. We would also like to thank two anonymous peer reviewers from the *Journal of Philosophical Transactions*, who commented on an earlier draft of this paper.

# 7.REFERENCES